\documentstyle[12pt,epsf]{article}

\begin{document}
\setlength{\oddsidemargin}{-1cm}
\setlength{\evensidemargin}{-1cm}
\setlength{\unitlength}{1cm}
\begin{flushright}
\normalsize
Freiburg-THEP 97/20\\
September 1997\vspace{0.8cm}\\
\end{flushright}

\begin{center}
{\Large\bf NLO correction to Higgs boson parameters in the 1/N expansion}\vspace{1.2cm}\\
T. Binoth\footnote{Address after November 1, 1997: 
                   Laboratoire de Physique Theorique ENSLAPP,
                   LAPP, B.P. 110, F--74941 Annecy--le--Vieux, France }, 
A. Ghinculov\footnote{Address after October 1, 1997: 
                       Randall Laboratory of Physics, University of Michigan, 
		       Ann Arbor, Michigan 48109--1120, USA}, 
J.J. van der Bij \vspace{0.4cm}\\
\it Albert-Ludwigs-Universit\"at Freiburg\\
\it Fakult\"at f\"ur Physik\\
\it Hermann-Herder-Stra\ss e 3\\
\it D-79104 Freiburg\vspace{1.2cm}\\
\end{center}

\begin{center}
{\bf Abstract}\vspace{0.4cm}\\
\begin{minipage}{16cm}
We present the result of a calculation of the next-to-leading correction to the Higgs propagator in
the 1/N expansion, where the Higgs sector is treated as an O(N) symmetric
$\sigma$-model. The results are compared with two-loop
perturbation theory.
The existing discrepancy between the lowest order of the 1/N expansion and perturbation theory
is dramatically reduced by including the NLO in 1/N.
We find a maximum effective Higgs mass of 930-980 GeV. We give an
approximate relation between Higgs width and mass, which can be used for
phenomenological purposes.
\end{minipage}
\vspace{0.8cm}\\
\end{center}
With the discovery of the vector bosons behind us, the next major question
in the phenomenology of the weak interactions is the search for the Higgs boson. Within the standard model the properties of the Higgs boson 
are determined when its mass is fixed, as the tree level selfcouplings 
are proportional to $m_H^2$. As long as the Higgs particle is light
there are therefore no fundamental problems in determining its properties from perturbation theory. This perturbation theory has recently even been extended
to the two-loop level [1-3]. However, when the Higgs particle becomes heavy (O(TeV))
the selfcoupling becomes large and perturbation theory becomes unreliable.

Therefore one would like to find an approximation to Higgs physics which
is applicable beyond perturbation theory. When one realizes that the Higgs sector of the standard model is nothing but an O(4) linear sigma-model, the
expansion in 1/N of the O(N)-symmetric sigma model suggests itself.
The 1/N expansion has a long history [4-13] and has also been applied
to the Higgs boson of the standard model [14,15]. The results in lowest order  are interesting and indicate an upper mass to the Higgs boson, with a non-perturbative width. However the situation is not satisfactory as the lowest order is very far from perturbation theory for small Higgs mass. 
This discrepancy has led to some doubts as to the usefulness of the 
1/N expansion. In order to clarify the situation it is therefore clearly desirable to know the next-to-leading order contribution of O(1/N). 

Although a large amount of work has been
done on the O(N) model, most of it is limited to the leading order, $O(1/N^0$).
Root [7,8] wrote down expressions in higher order, but did not explicitly evaluate them. The reason for this is twofold. First of all the expressions are quite complicated. More fundamental is the presence of a 
tachyon in the Higgs propagator. Although some of the early papers [9-11] suggested that this
problem could be solved, later work [12,13] showed these claims to be unfounded.
For practical calculations some prescription of dealing with the tachyon 
is needed. Our philosophy is based on the observation that the tachyon is beyond the Landau pole, which 
is located at a non-perturbative scale $\mu e^{48\pi^2/\lambda}$ and corresponds
to physics that we do not understand well. Our goal here is to find
a resummed expression  for the Higgs propagator, which after expansion in $\lambda$ should reproduce perturbation theory. The effects of the tachyon, being non-perturbative, should not contribute. The exact behaviour at the location of the tachyon pole is expected to have little effect on the Higgs
lineshape, which is in a different region of s. Therefore, whenever we have a propagator with a tachyon pole present inside a graph, we  simply subtract the pole part at the location of the tachyon.  With this prescription the calculation
becomes well defined, and one can represent the O(1/N) contribution in terms of a number of Feynman-like diagrams.

The starting point of the calculation is the following Lagrangian:
\begin{equation}
{\cal L} = \frac{1}{2} \partial_{\mu}\phi^i \partial_{\mu}\phi^i 
-\frac{1}{2} \mu^2 \phi^i \phi^i - \frac{\lambda}{4!N} (\phi^i \phi^i)^2 
\end{equation}
                          
Here $\phi^i$ is a scalar field with N components $i=1, ...,N$.
In principle one could use this Lagrangian in order to calculate 
contributions in different orders of 1/N. However the combinatorics becomes
very complicated and it is advantageous to introduce an auxiliary field $\chi$,
and add a nondynamical term to the Lagrangian as follows \cite{cjp}:
\begin{equation}
 {\cal L} = {\cal L} + \frac {3 N}{2 \lambda} (\chi - \frac{\lambda}{6 N}
\phi^i \phi^i - \mu^2)^2 =                                           
 \frac{1}{2} \partial_{\mu}\phi^i \partial_{\mu}\phi^i + \frac{3 N}{2 \lambda} \chi^2
 -\frac{1}{2} \chi \phi^i \phi^i -\frac {3 \mu^2 N}{\lambda} \chi + constant 
\end{equation}
 
This form of the Lagrangian has the same physical content as before, however the Feynman rules are changed, thereby facilitating the counting of degrees
in 1/N. For doing the higher order calculation the introduction of the extra
field $\chi$ is practically the only way not to get lost in the combinatorics.
In order to perform the 1/N correction to the Higgs-propagator, we will have 
to renormalize the Lagrangian. One should be careful to insure that one is in the broken phase and that the Goldstone theorem is valid.
The details are to technical to be presented here. We only wish to note that
the Higgs v.e.v. scales as $\sqrt{N} v$, so for comparison with experiment
one should take N=4 and $v = 123 GeV$.

The diagrams of leading order in 1/N change the propagators
of $\sigma, \chi$ to propagators containing a bubble sum of internal
Goldstone-loops and one finds the following propagators [14,15]:
\begin{eqnarray}
D_{\sigma \sigma} &=& \frac{i(3/\lambda + 1/2 B(s))}
{s(3/\lambda +1/2 B(s)) - v^2} \cr
D_{\chi \chi} &=& \frac{1}{N}  \frac{i s}
{s(3/\lambda +1/2 B(s)) - v^2} \cr
D_{\chi \sigma} &=& \frac{v}{\sqrt{N}}  \frac{i}
{s(3/\lambda +1/2 B(s)) - v^2} \cr
D^{ij}_{\pi \pi} &=& \frac{i \delta^{ij}}{s}
\end{eqnarray}

Here B(s) is the basic bubble integral, which after subtraction of 
a pole in $(n-4)$ and defining a renormalization scale $\mu$ is given by:
\begin{equation}
B(s) = \frac {1}{16 \pi^2} ln( \frac {\mu^2}{-s - i \epsilon} )    
\end{equation}

With these propagators defined it is now straightforward to write diagrams  
 for the next-to-leading order contribution
O(1/N) to the Higgs-propagator. The vertices can be directly read  
from the Lagrangian of eq.(2). The propagators are the expressions given above,
where one should still take care to subtract the pole at the location of the tachyon.
Because of the complicated propagators the evaluation of the graphs
is highly involved and can only be performed numerically. The graphs also
involve overall divergences and subdivergences, wich were handled by
making subtractions at intermediate renormalization scales. In order to do the
actual calculation new techniques were necessary, extending the methods of ref.[16]. Details of the calculation, including renormalization and tachyon subtraction,  will be given in a subsequent publication.

We note that we need to evaluate the two-point functions for $\sigma \sigma,
\chi \chi, \chi \sigma$. After this we have to invert the matrix in order
to project out the physical Higgs propagator. This final Higgs propagator
is then a physical quantity, which can in principle be measured in the process
$\mu \bar \mu \rightarrow t \bar t$, after taking into account the Yukawa coupling renormalization. The resulting lineshapes can be compared with the
known perturbative result [17]. The two-loop and the NLO 1/N lineshapes
are so close together that it is not very enlightening to plot examples. 

\begin{figure}
\hspace{1.cm} 
    \epsfxsize = 14.5cm
    \epsffile{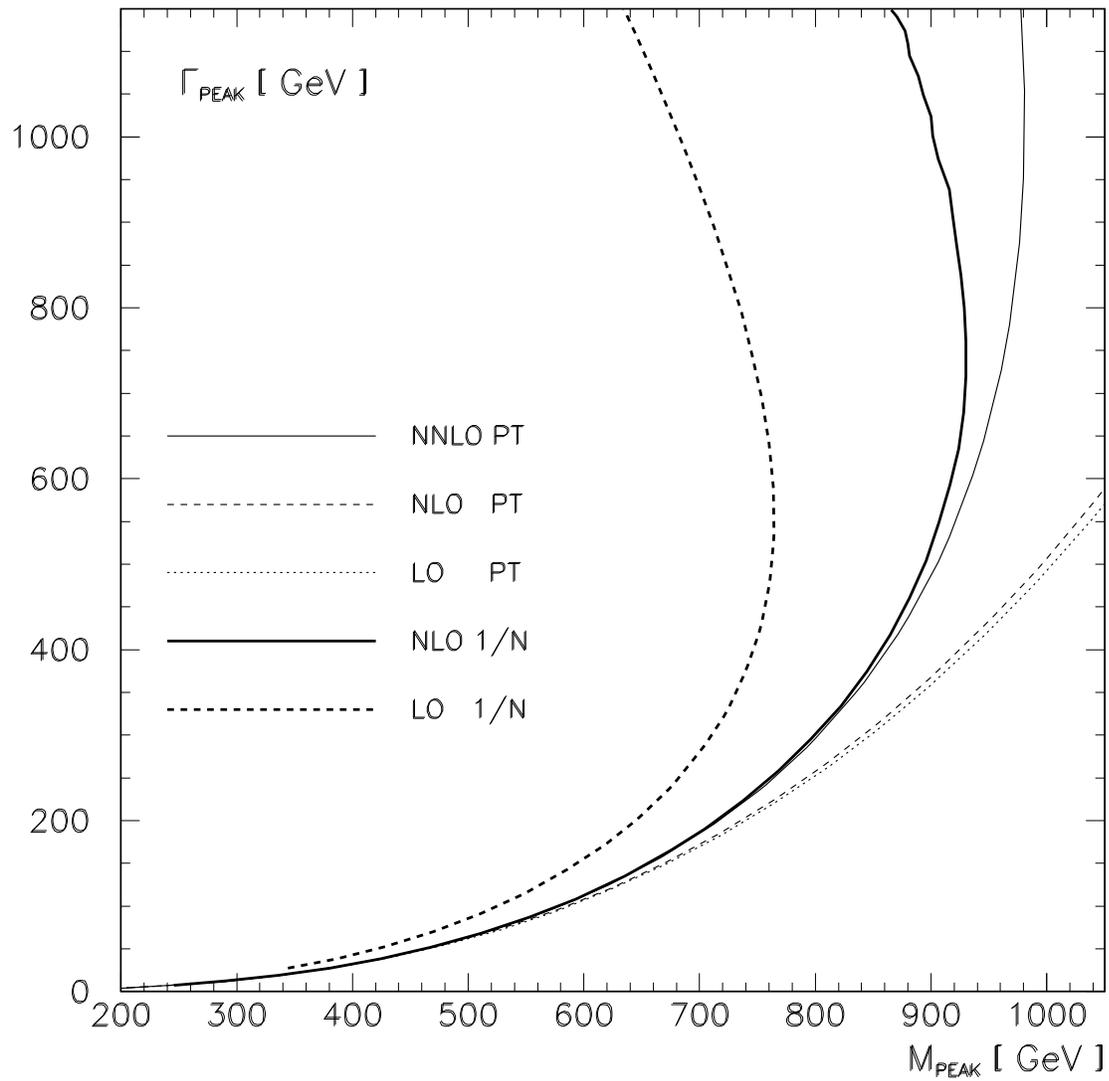}
\caption{{\em Width versus mass of the Higgs-boson 
              in perturbation theory and in the 1/N expansion.}}
\end{figure}

Instead, to study the results in detail, we plotted in fig.(1)
an effective Higgs width against the Higgs mass. Ideally one would like to plot
here the real part and the imaginary part of the pole of the Higgs propagator.
However our program does not allow us at the moment to calculate the graphs
for complex momentum. We therefore define the Higgs mass $m_{peak}$ as the location 
of the peak of the cross section. The width $\Gamma_{peak}$ is taken from the
peak height of the cross section as if the line-shape was a Breit-Wigner
function:
\begin{equation}
\sigma (s) \sim \frac {1}{(s-m_{peak}^2)^2+m^2_{peak}
\Gamma^2_{peak}}
\end{equation}

Of course the lineshape is not exactly of Breit-Wigner type, however we found
that this definition of the width is within a few percent of for instance
the width at half-maximum. This parametrisation is therefore a good measure of the leading features of the Higgs lineshape.

We notice from the figure that the 1/N expansion and perturbation theory
appear to be converging towards a common relation between the Higgs width and mass. Even for Higgs masses as large as 900 GeV, the agreement between the
two-loop result and the 1/N expansion is spectacularly close. Only when 
one comes close to the maximum Higgs mass do differences occur. These differences can easily be due to the neglect of higher order terms in the 
expansions. We therefore trust that we have reached a correct understanding of 
the Higgs lineshape, the remaining uncertainties being given by the area 
between the NLO in 1/N and the two-loop curves. The maximum
possible Higgs mass $m_{peak}$ is between 930 GeV and 980 GeV.

The results given above can strictly speaking not be used directly at the LHC,
because one should also take higher order 1/N corrections in the Higgs-Goldstone
boson vertex into account. However we expect, that the correction to the width will be the dominant feature in practice.
We notice that the Higgs width near the maximum mass is significantly larger
than the tree level or one-loop width. This will effect the measured lineshapes at the LHC and might change the discovery limits. When the Higgs
becomes wide the signal becomes washed out and lost in the background.
For phenomenological purposes we therefore give here an approximate 
relation between the Higgs mass and the Higgs width. Expressing $m_{peak}$ and $\Gamma_{peak}$ in TeV, one can use the following approximate relation:

\begin{equation}
  M_{peak}^3  =    2.03  \cdot \Gamma_{peak}   
	           - 0.81 \cdot \Gamma_{peak}^2 
                   - 1.02  \cdot \Gamma_{peak}^3 
	           + 0.54 \cdot \Gamma_{peak}^4  
		~~ ,
\end{equation}
which is a fit to the NLO $1/N$ curve of fig. 1.
This can be compared with the perturbative result:

\begin{equation}
  M_{peak}^3  =    2.09  \cdot \Gamma_{peak}   
	           - 1.24 \cdot \Gamma_{peak}^2 
                   - 0.02  \cdot \Gamma_{peak}^3 
	           + 0.12 \cdot \Gamma_{peak}^4  
		~~ .
\end{equation}

\noindent {\bf Acknowledgements} We acknowledge useful discussions with 
Dr. G. Jikia and Dr. B. Kastening. The work of A.G. was supported by the Deutsche Forschungsgemeinschaft (DFG).

\end{document}